**Trends, Impact, and Thematic Evolution of *Physics Education*: A Bibliometric and Topic Modelling Analysis of Six Decades of Publications**

Purwoko Haryadi Santoso[1*], Nurlina[1], Mutmainna[1,2]

[1] Department of Physics Education, Universitas Sulawesi Barat, Majene 91413, Indonesia

[2] Department of Educational Research and Evaluation, Universitas Negeri Yogyakarta, Sleman 55285, Indonesia

**Abstract**

*Physics Education* has played a prominent role in advancing physics education research since its establishment in 1966. However, a comprehensive investigation of its publication trends, impact, and thematic evolution over six decades remains limited. This study examines the development of *Physics Education* through a bibliometric analysis of publications indexed in Scopus database between 1966 and 2026. Following a PRISMA-based screening process, 6,311 documents were retained from an initial dataset of 7,709 records. Publication trends, citation impact, leading countries and institutions, topic modeling, and topic evolution were analyzed to examine the journal's development and thematic structure. The results indicate sustained growth in publication output, significantly after the digital and information era. The United Kingdom and the United States emerged as the most productive contributors and accumulated the highest total citations, whereas Germany and Norway achieved the highest citation impact in terms of citations per paper. This pattern was further reinforced by increasing contributions from countries in Asia and the Global South, reflecting the growing internationalization of the journal. Major research themes include force and motion experiments, energy and light experiments, laboratory-based learning, conceptual understanding, and physics teacher education. Topic prevalence analysis indicates that, while experimental and laboratory-based studies have remained a defining feature of *Physics Education*, increasing attention has been devoted to conceptual understanding, assessment, technology-supported instruction, and pedagogical innovation in recent decades. Overall, *Physics Education* has evolved from a journal primarily centered on laboratory experiments and instructional demonstrations into a multidisciplinary venue emphasizing conceptual understanding, educational technology, and contemporary innovations in physics teaching and learning.

## 1. Introduction

Since its establishment in 1966, *Physics Education* has become one of the most influential journals in the field of physics education research (PER), serving as a major platform for disseminating innovations in physics teaching, laboratory practices, curriculum development, assessment, and educational technologies. The journal has documented significant transformations in physics education, ranging from traditional laboratory demonstrations and conceptual learning approaches to contemporary themes such as STEM education, digital experimentation, computational tools, and quantum physics education. As a result, *Physics Education* represents a valuable source for understanding the intellectual structure of PER over an extended period.

Over the past several decades, PER has evolved into a mature interdisciplinary research domain that integrates perspectives from physics, cognitive science, educational psychology, and learning sciences (Docktor and Mestre 2014, McDermott and Redish 1999). Earlier research primarily focused on conceptual understanding, misconceptions, and problem-solving abilities, which later expanded toward inquiry-based learning, scientific practices, and evidence-based instructional strategies (Odden *et al* 2020, Yun 2020, Santoso *et al* 2022). More recently, the rapid development of educational technologies has stimulated research on smartphone-assisted laboratories (Zhao 2026, Kaps *et al* 2021, Kaps and Stallmach 2022), Arduino-based instrumentation (Organtini and Tufino 2022, Liu *et al* 2025), virtual learning environments, learning analytics, and emerging applications of artificial intelligence in physics education (Chen *et al* 2020, Pace *et al* 2024, Campbell *et al* 2024). These developments indicate that the research landscape of physics education is becoming increasingly diverse and technologically oriented.

The growing maturity of PER has been accompanied by a substantial increase in publication volume. Consequently, bibliometric studies have become an important approach for identifying research trends, influential contributors, intellectual structures, and future research directions. In parallel, several review studies have synthesized the accumulated knowledge within the field. For example, Docktor and Mestre (2014) reviewed PER literature and identified major research areas, including conceptual understanding, problem solving, assessment, cognitive psychology, attitudes and beliefs. They also highlighted several research gaps and directions for future studies. These studies have provided valuable insights into the development of PER and have helped clarify both established research traditions and emerging areas requiring further investigation.

Despite these contributions, several important gaps remain. Former PER scholars have provided valuable insights into the field. Odden et al (2020) examined publications from the *Physics Education Research Conference* (PERC) and identified major intellectual trends within the PER community using natural language processing (NLP) technique. Yun (2020) analyzed research themes in the *American Journal of Physics* (AJP) and *Physical Review Physics Education Research* (PRPER), while Santoso et al (2022) mapped the thematic landscape of physics education studies presented at some Indonesian international conferences on science education. Although these thematic studies have expanded our understanding of PER, they still focus primarily on selected journals, conference proceedings, or regional publication venues. As a result, they might provide only a partial view of the intellectual development in the field.

To the best of our knowledge, the long-term evolution of *Physics Education*, one of the oldest and most influential journals in physics education research, remains largely unexplored. Established in 1966, *Physics Education* has continuously documented major transformations in physics teaching and learning, ranging from traditional laboratory demonstrations and conceptual learning to contemporary topics such as STEM education, smartphone-based experimentation, digital laboratories, and quantum physics education. Unlike previous studies, which concentrated on specific segments of the PER community, a comprehensive investigation of how publication productivity, citation impact, geographical participation, institutional contributions, and research themes have evolved within *Physics Education* over six decades is still lacking. Consequently, there is limited understanding of the journal's role in shaping the historical development and intellectual structure of physics education research. Finally, the emergence of new contributors from Asia, Latin America, and other Global South regions suggests a changing geographical landscape in PER, yet the extent of this transformation has not been systematically investigated within the context of *Physics Education*.

Addressing these gaps is important for both the PER community and educational researchers. As one of the longest-running journals in the discipline, *Physics Education* provides a unique lens through which to examine how research priorities, influential contributors, and thematic structures have evolved across generations of physics educators and researchers. A comprehensive bibliometric assessment can therefore offer a deeper understanding of the journal's role in shaping the development of PER and reveal emerging directions that may influence future research agendas. The present study has conducted the first comprehensive bibliometric analysis of *Physics Education* spanning six decades of publication (1966-2026). By integrating publication trends, country and institutional contributions, citation impact, topic modeling, and topic evolution analysis, this study provides a contribution to the journal's development and identifies emerging directions that may shape the future of PER.

Using a bibliometric dataset of publications indexed in Scopus database over six decades (1966-2026), the present study combines productivity and citation analyses with topic modeling and topic evolution analysis to provide a comprehensive overview of the journal's historical development and intellectual structure over time. The study is guided by the following research questions:

1. How has the publication productivity of *Physics Education* evolved over six decades?
2. Which countries and institutions have contributed most significantly to the journal?
3. Which countries and institutions have achieved the highest citation and impact?
4. What are the major thematic structures of *Physics Education* research?
5. How have research themes evolved from 1966 to 2026, and which topics are emerging as future directions within the physics education community?

Accordingly, this study contributes to the literature by presenting the first comprehensive bibliometric analysis of *Physics Education* spanning six decades of publication (1966-2026). By integrating publication performance, geographical and institutional contributions, citation impact, and topic modeling, the study adds a holistic understanding of the journal's historical development, intellectual structure, and emerging research directions, thereby providing valuable insights into the past, present, and future of physics education research community.

## 2. Methods

### 2.1 Research Design

This study employed a bibliometric research design to examine the publication productivity, citation impact, topic modelling, and thematic evolution of *Physics Education* from 1966 to 2026. Bibliometric analysis is widely recognized as an effective approach for investigating the development of scientific fields through the quantitative examination of publication metadata, citation patterns, collaboration structures, and conceptual knowledge domains (Aria and Cuccurullo 2017). By systematically analyzing bibliometric indicators, the methods can reveal the historical development, influential contributors, and emerging research trends within a scientific discipline.

To address RQ1, publication performance analysis was conducted to investigate the growth of publication productivity over six decades, including annual publication output and long-term publication trends. To answer RQ2, geographical and institutional contribution analyses were performed to identify the countries and institutions that contributed most significantly to the journal based on publication output. For RQ3, citation impact analysis was carried out using citation-based indicators, including total citations (TC), citations per paper (CPP), and h-index, to identify countries and institutions with the highest scholarly influence in the field.

To address RQ4 and RQ5, the intellectual structure of the journal was examined through topic modeling using Latent Dirichlet Allocation (LDA) based on the abstracts of published papers. Topic modeling was employed to extract the major themes represented in the journal, while temporal topic analysis was used to examine how research themes evolved between 1966 and 2026. Furthermore, topic prevalence across publication periods was analyzed to identify increasing, stable, and declining themes, thereby providing insights into the long-term development of research interests and potential future directions within the PER community.

Accordingly, the research framework consisted of four complementary stages: (1) publication performance analysis, (2) geographical and institutional contribution analysis, (3) geographical and institutional citation impact analysis, and (4) thematic analysis through topic modeling using LDA analysis.

### 2.2 Data Source

The bibliographic dataset was retrieved from the Scopus database. Scopus was selected because it is one of the largest multidisciplinary citation databases and provides comprehensive metadata suitable for bibliometric analysis, including author names, publication year, author affiliations, article titles, keywords, abstracts, and citation records. The Scopus search was restricted to publications indexed in *Physics Education* from 1966 to 9 August 2026, the date on which the dataset was retrieved.

Our search was exported in BibTeX (.bib) format to facilitate a more complete database. The file was then imported and processed using the *bibliometrix* package in R, enabling the extraction, organization, and analysis of publication metadata. These records formed the primary dataset for subsequent screening, cleaning, and further analyses.

**2.3 Exclusion and Inclusion Criteria**

A PRISMA (*Preferred Reporting Items for Systematic Reviews and Meta-Analyses*)-based screening procedure was employed to ensure the quality and consistency of the bibliometric dataset. The procedure provided a transparent framework for record identification, screening, eligibility assessment, and final inclusion prior to bibliometric analysis. The initial search in the Scopus database retrieved 7,709 records published in *Physics Education* between 1966 and 2026.

To ensure the quality and consistency of the dataset, explicit inclusion and exclusion criteria were established. The inclusion criteria comprised publications classified as Article, Letter, Note, Review, and Short Survey, as these document types represent peer-reviewed scholarly contributions suitable for bibliometric analysis. In addition, only records containing reported author affiliations and complete abstracts were retained, particularly to support thematic analysis and topic modeling.

Based on these criteria, records that did not meet the above requirements were excluded, specifically those outside the selected document types (e.g., Conference Paper, Editorial, and Erratum, which do not represent original research contributions comparable to the included types) and those falling short of the metadata completeness criteria described above, in order to ensure the reliability of the institutional, geographical, and thematic analyses. Duplicate records were also identified and removed to avoid redundancy in the dataset.

A summary of the screening and selection process is presented in Fig. 1. Following the PRISMA-based procedure, 7,709 records were initially identified in Scopus. In the first screening stage, duplicate records were identified and removed to avoid redundancy and the potential overestimation of publication and citation indicators. A total of 86 duplicate records were eliminated, reducing the dataset from 7,709 to 7,623 unique publications.

Duplicate detection was performed using the *bibliometrix* package in R, which facilitated the systematic identification and removal of redundant records within the Scopus dataset.

In the second stage, non-research document types were excluded to ensure that the dataset represented original scholarly contributions. Conference papers, editorials, and erratum were removed because these records generally function merely as communication, editorial, or correction materials rather than primary research outputs. This screening step excluded 87 records, reducing the dataset from 7,623 to 7,536 publications for the subsequent eligibility assessment.

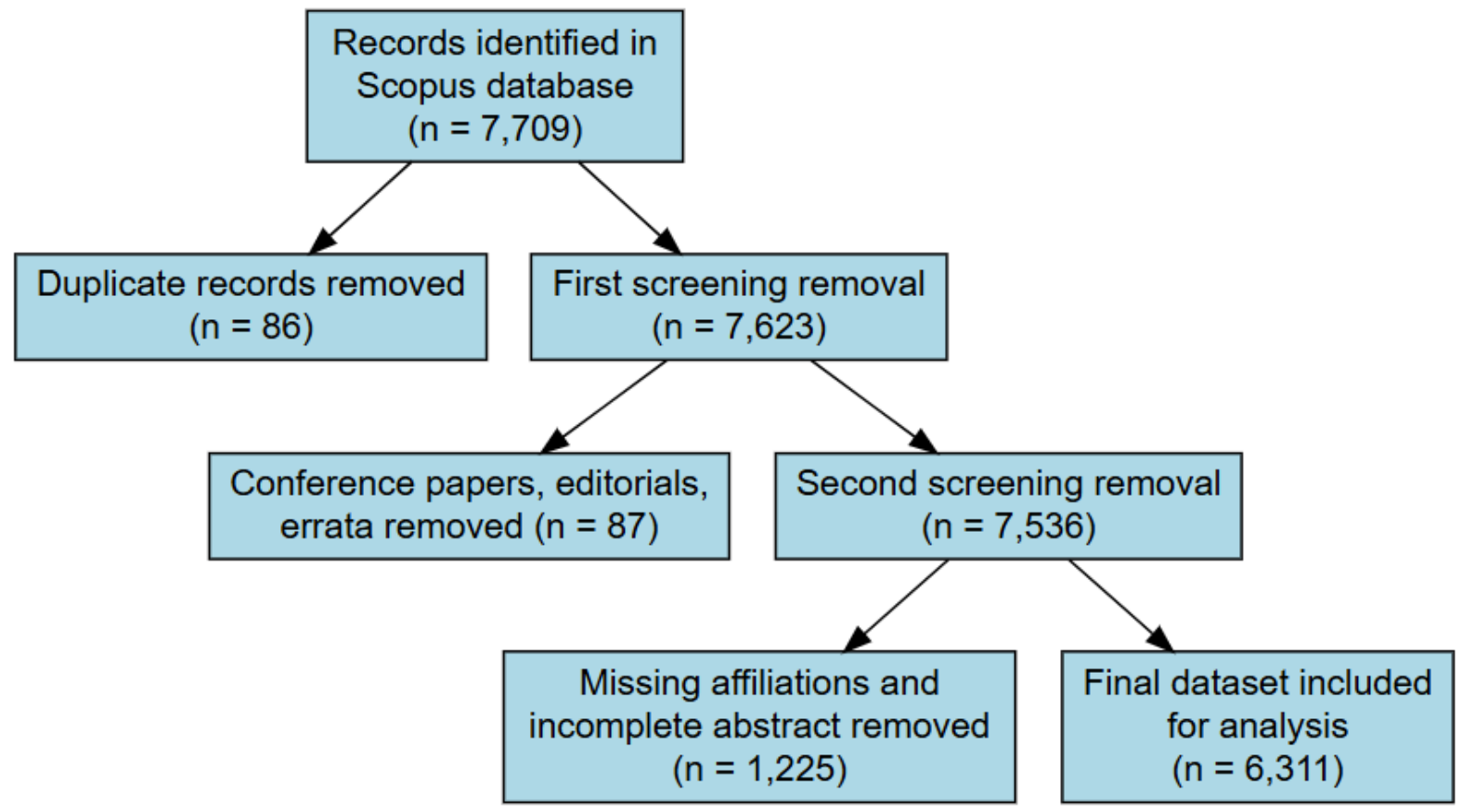

**Fig.1** Screening Results based on PRISMA Guideline

In the third screening stage, publications with missing affiliation information and incomplete abstracts were excluded. Institutional affiliation data are essential for country-, institution-, and collaboration-level analyses, whereas abstracts constitute the primary textual source for the topic modeling procedures employed in this study. Records without affiliation information could not be reliably assigned to institutions or countries and were therefore unsuitable for bibliometric mapping. Similarly, publications lacking complete abstracts could not be incorporated into the need for thematic analysis. As a result, 1,225 records were excluded at this stage, resulting in a final corpus of 6,311 publications for subsequent bibliometric and topic modeling analyses.

Overall, the screening process excluded 1,398 records (18.1%) from the initial dataset, resulting in a final corpus consisting exclusively of unique research publications with complete affiliation information. Nevertheless, missing affiliation data were observed throughout the entire publication period from 1966 to 2026. The proportion of records with unreported affiliations varied across publication years, averaging approximately 18-20% per annual publication (Fig. 2). Particularly high proportions were observed during the early 1980s and early 1990s, when the percentage of publications with unreported affiliation information sometimes exceeded 40% per year. This issue should be considered when interpreting the findings because the exclusion of records with missing affiliations may influence estimates of country- and institution-level productivity. Furthermore, the reduction in the number of eligible records may introduce some bias into the topic modeling analysis, particularly for periods in which missing affiliation data were more prevalent.

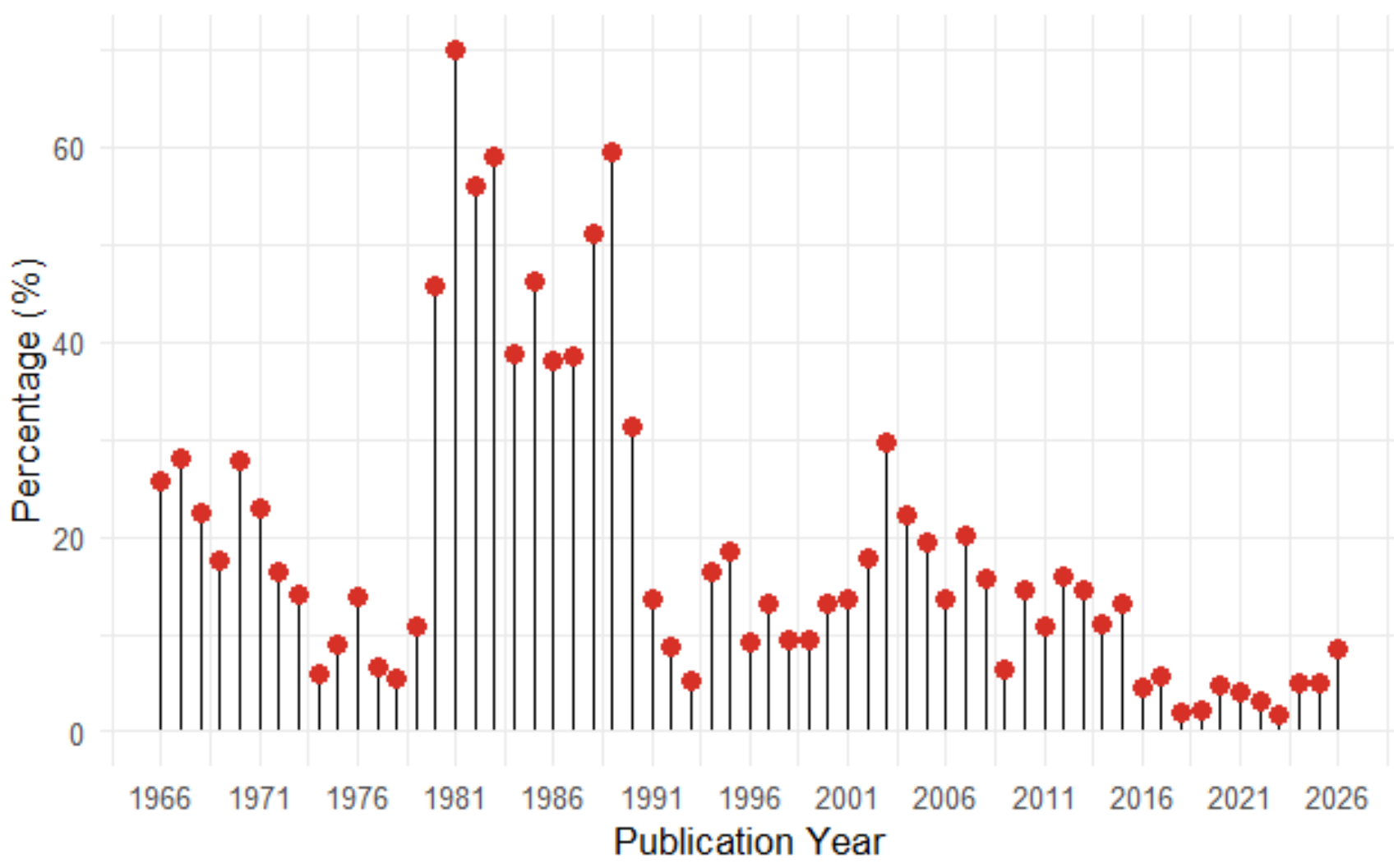

**Fig. 2** Annual Percentage of Records with Unreported Affiliations

The retained records were subsequently subjected to data cleaning and standardization procedures. Variations in country names (e.g., *USA* versus *United States* and *UK* versus *United Kingdom*) were harmonized, and institutional names were standardized to mitigate inconsistencies caused by alternative affiliation formats. Records containing incomplete or ambiguous affiliation information were manually reviewed and corrected whenever possible to enhance data quality and ensure the robustness of the interpretations. The resulting dataset was then prepared for statistical analysis**.** The cleaned dataset was subsequently saved in CSV format and used for all bibliometric and thematic analyses (topic modeling).

### 2.4 Bibliometric Analysis

To address RQ1-RQ3, a set of descriptive bibliometric indicators was employed to assess publication productivity, scientific contribution, and citation impact. For RQ1, publication performance analysis was conducted using annual publication counts to examine the growth and evolution of *Physics Education* over six decades. It should be noted that publications with missing affiliation information were excluded only from the country-, institution-, and topic-level analyses. The publication trend analysis presented in RQ1 was based on all records retained after the second screening stage ($n$ = 7,536).

For RQ2, contribution analyses were performed using the final dataset ($n$ = 6,311) to identify the most productive countries and institutions. Publication records were aggregated based on affiliations, and productivity indicators were calculated using the total number of publications contributed by each country and institution. For institution- and country-level analyses, each publication was assigned to the primary affiliation of the first author as reported in the Scopus database. Therefore, publications involving multiple authors from different institutions or countries were counted only once, based on the first author's affiliation, to avoid double counting and ensure consistent attribution of research output.

For RQ3, publication impact analysis was conducted using some citation-based indicators, including total citations (TC), citations per paper (CPP), and h-index. TC was used to measure the overall citation influence accumulated by a country or institution, whereas CPP assessed the average citation impact per publication. The h-index was included as a complementary measure that combines productivity and citation performance by identifying the number of publications that received at least an equivalent number of citations. Together, these indicators enabled the identification of countries and institutions with the highest scientific influence and research impact within the journal. Comparative analyses of publication output and citation performance further distinguished highly productive contributors from highly influential contributors. Comparative analyses using visualization facilitated by *ggplot* package in R were conducted to identify whether highly productive countries and institutions also demonstrated high citation influence.

### 2.5 Topic Modeling Analysis

To extract the major research themes (RQ4 and RQ5), abstract texts were subjected to topic modeling. The method basically followed our former study (Santoso *et al* 2022). Prior to topic modeling, abstracts were converted into a corpus and preprocessed through stop-word removal, elimination of domain-independent terms, and lemmatization. A document-term matrix was then generated, retaining only terms with a minimum length of three characters. To improve computational efficiency and reduce noise, sparse terms were removed using a sparsity threshold of 0.99 before conducting the LDA analysis.

Latent Dirichlet allocation (LDA), one of the natural language processing (NLP) methods, was employed to identify latent thematic structures within the corpus of article abstracts. Broadly speaking, LDA assumes that each document is characterized by a mixture of topics and that each topic is represented by a probability distribution over words (Blei *et al* 2003). Through this probabilistic framework, the model identifies groups of terms that frequently co-occur across documents (document-term matrix) and uses these patterns (corpus) to infer underlying research themes in the analyzed texts. As a result, publications discussing similar concepts can be clustered into common topical domains without requiring predefined thematic categories.

To determine the optimal number of topics ($K$), a series of LDA models with varying topic numbers were estimated and compared. Model selection was guided by a combination of statistical fit indices, topic coherence, and semantic interpretability (Röder *et al* 2015). The objective was to identify a topic solution that maximized thematic distinctiveness while minimizing overlap among topics. Based on this evaluation process and qualitative interpretation by the authors, the five-topic solution ($K = 5$) was determined as the most appropriate representation of the thematic structure of *Physics Education*.

Following model estimation, each publication was assigned topic probabilities indicating its degree of association with each latent topic. Topic interpretation was conducted by examining the highest-probability keywords, representative articles, and thematic content associated with each topic. The resulting topic labels were concluded through researcher interpretation rather than directly by the algorithm. The five identified topics were subsequently used as the basis for analyzing thematic prevalence, thematic evolution, and long-term research trends throughout the history of the journal. To investigate thematic changes over time, publications were divided into three periods, and topic changes were examined at five-year intervals. Topic prevalence was analyzed longitudinally to identify thematic trajectories across the journal's history.

## 3. Results

### A. Bibliometric Analysis

Fig. 3 illustrates the publication growth of *Physics Education* between 1966 and 2026. Three distinct developmental phases can be identified. The early growth phase (1966-1981) was characterized by a steady increase in publication output, rising from fewer than 80 publications per year to more than 120 publications annually by the late 1970s. This growth was followed by a stagnation phase (1981-2001), during which publication output remained relatively stable, fluctuating around 80-100 publications per year. The stability observed during this period may reflect the consolidation and maturation of the PER community. The most substantial expansion occurred during the digital and information era (2001-2026). Following the widespread adoption of the internet and the increasing availability of online scholarly publishing, publication output grew rapidly and consistently. Improved access to scientific knowledge, digital dissemination of research, and broader international participation likely contributed to this growth.

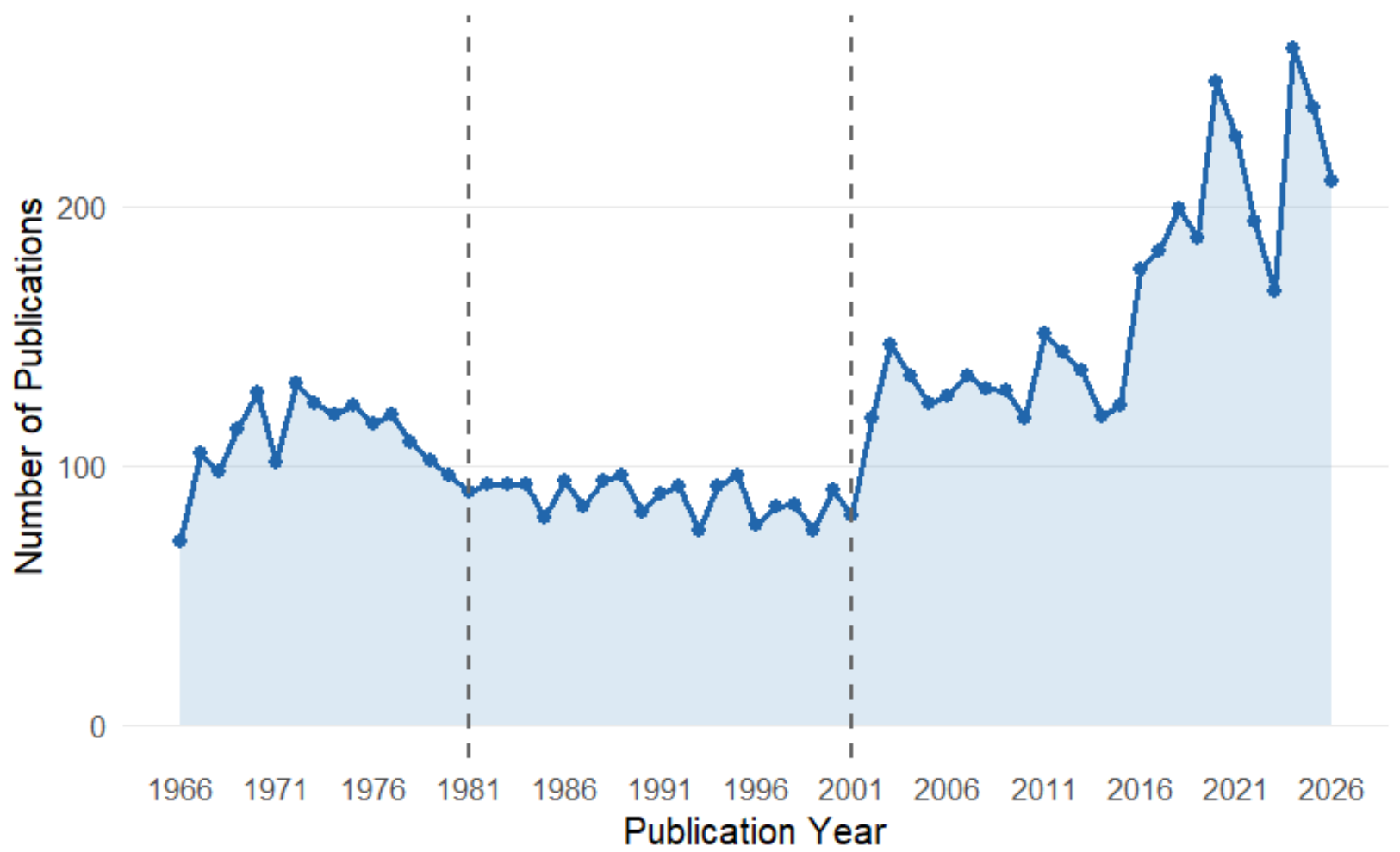

**Fig. 3** Number of Publications in *Physics Education* Between 1966 and 2026

Fig. 4 illustrates the 25 most productive countries (A) and institutions (B) contributing to *Physics Education*. Although publication output was strongly dominated by the United Kingdom (2,654) and, to a lesser extent, the United States (518) and Australia (283), the results also reveal the emergence of several countries from Asia and the Global South. Notable examples include Brazil (177), India (162), Turkey (112), Indonesia (95), Israel (90), Singapore (81), Thailand (69), Hong Kong (68), Japan (60), China (60), and Chile (43). The presence of these countries among the most productive contributors suggests that PER is no longer concentrated exclusively within traditional research centers in Europe and North America. Instead, scholarly contributions increasingly originate from diverse geographic contexts, reflecting the growing internationalization of the field.

At the institutional level, publication productivity was concentrated among a relatively small group of universities, led by the University of Sydney (129 publications), followed by King's College London (77), Science on Stage Europe (61), and the University of Cambridge (52). Notably, institutions from the United Kingdom accounted for a substantial proportion of the most productive organizations, mirroring the country's dominance at the country level. Hence, these results suggest that the development of physics education research published in the journal has been strongly shaped by a limited number of countries and institutions, while also showing increasing participation from emerging research communities in Asia and the Global South countries.

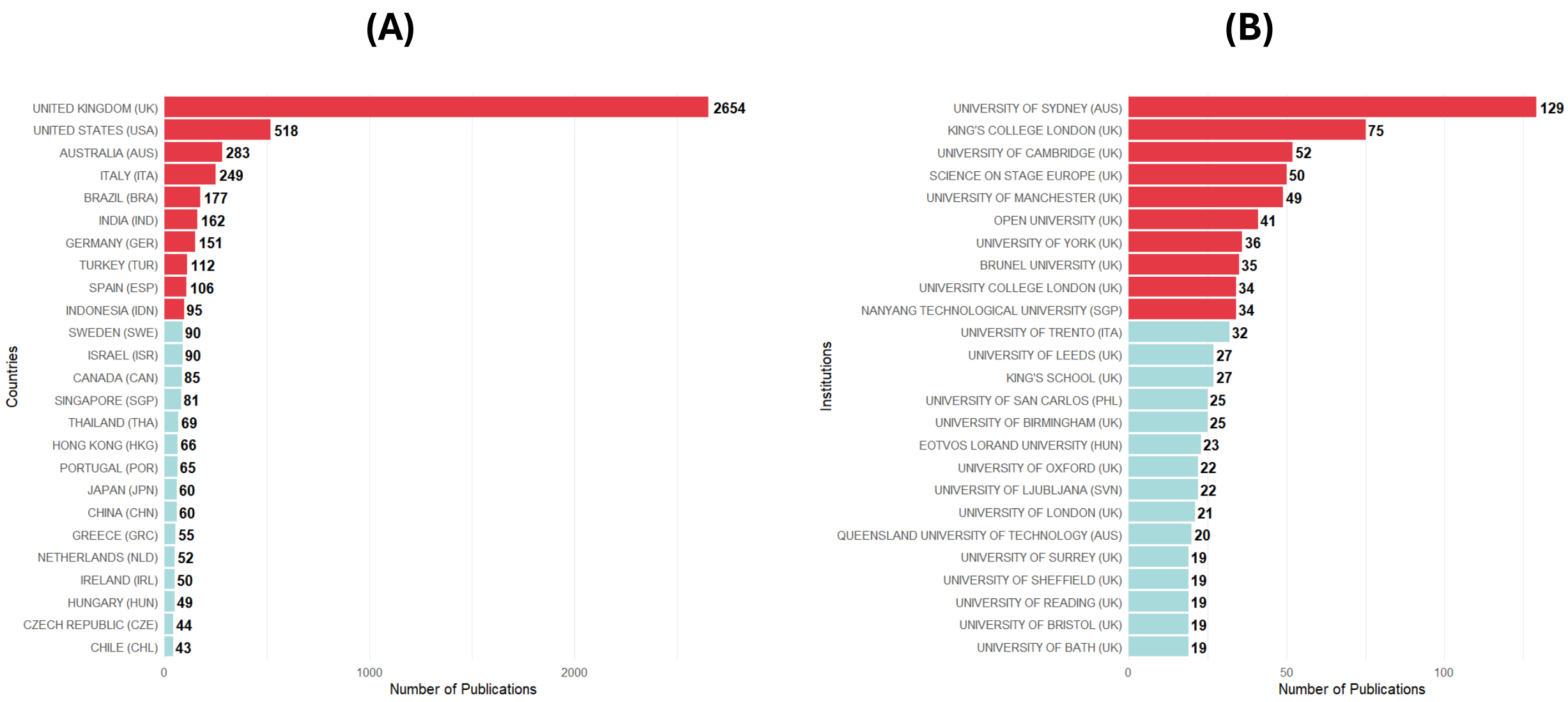

**Fig. 4** Most Productive Countries (A) and Institutions (B) in *Physics Education*

Fig. 5 presents the 25 most cited countries (A) and institutions (B) in *Physics Education*. The citation landscape largely mirrors the publication productivity patterns shown in Fig. 4, although notable differences remain. At the country level, the United Kingdom accumulated the highest number of citations (6,413), followed by the United States (2,072), Germany (1,567), Australia (1,365), and Italy (1,314). Other highly cited countries included Sweden (704), Brazil (629), Spain (579), Israel (523), and Portugal (521).

The results also highlight the increasing internationalization of *Physics Education*. Beyond the traditional leaders, Brazil (629 citations), Turkey (495), Indonesia (420), Singapore (400), and India (388) emerged among the most cited countries. Their strong citation performance demonstrates the growing visibility and influence of physics education research originating from emerging research communities outside the historical centers of physics education research (PER) in North America and Europe.

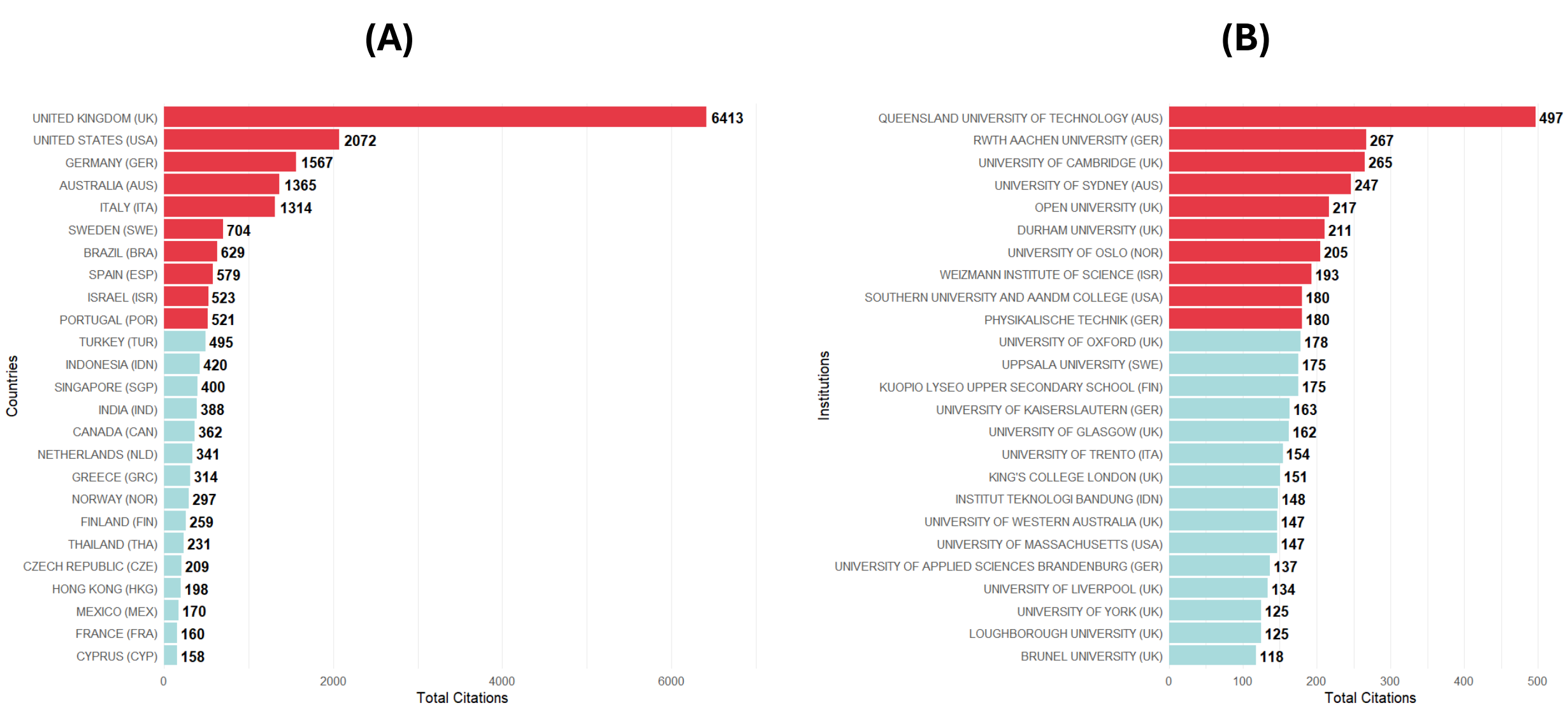

**Fig. 5** Most Cited Countries (A) and Institutions (B) in *Physics Education*

The institutional citation rankings further indicate that research influence is still concentrated within a relatively small number of highly visible institutions, although impactful contributions continue to emerge from diverse educational settings. Queensland University of Technology ranked first with 497 citations, substantially outperforming all other institutions. It was followed by RWTH Aachen University (267 citations), the University of Cambridge (265), the University of Sydney (247), and Uppsala University (231). Other highly cited institutions included Open University (217), Durham University (211), University of Oslo (205), and the Weizmann Institute of Science (193). The prominence of universities from the United Kingdom, Australia, Germany, and Scandinavia reflects the strong international influence of these countries within *Physics Education*. Interestingly, several non-university institutions, including Southern University and A&M College, Physikalische Technik, and Kuopio Upper Secondary School, also appeared among the most cited organizations. This finding suggests that influential contributions to *Physics*

*Education* extend beyond research-intensive universities and include schools and specialized educational institutions actively engaged in physics education innovation.

Fig. 6 presents the leading countries (A) and institutions (B) ranked by citation per paper (CPP) in the journal. The ranking differs substantially from the publication productivity and total citation patterns shown in Fig. 4 and Fig. 5. At the country level, Norway achieved the highest CPP (11.88), followed by Germany (10.38), Finland (10.36), and Austria (9.06). Portugal (8.02) and Sweden (7.82) also recorded relatively high citation impact. Notably, several countries that were among the most productive contributors, such as the United Kingdom and the United States, did not appear at the top 10 of the CPP ranking. This finding suggests that high publication volume does not necessarily translate into higher average citation impact per published article.

The results further indicate that CPP measure was distributed across both established and emerging research communities. In addition to several European countries, South Africa (6.89), Uruguay (6.09), Israel (5.81), Morocco (5.07), and Singapore (4.94) appeared among the leading countries by CPP. These findings suggest that publications originating from a number of countries outside Europe attracted citation rates comparable to those of several traditionally established research systems.

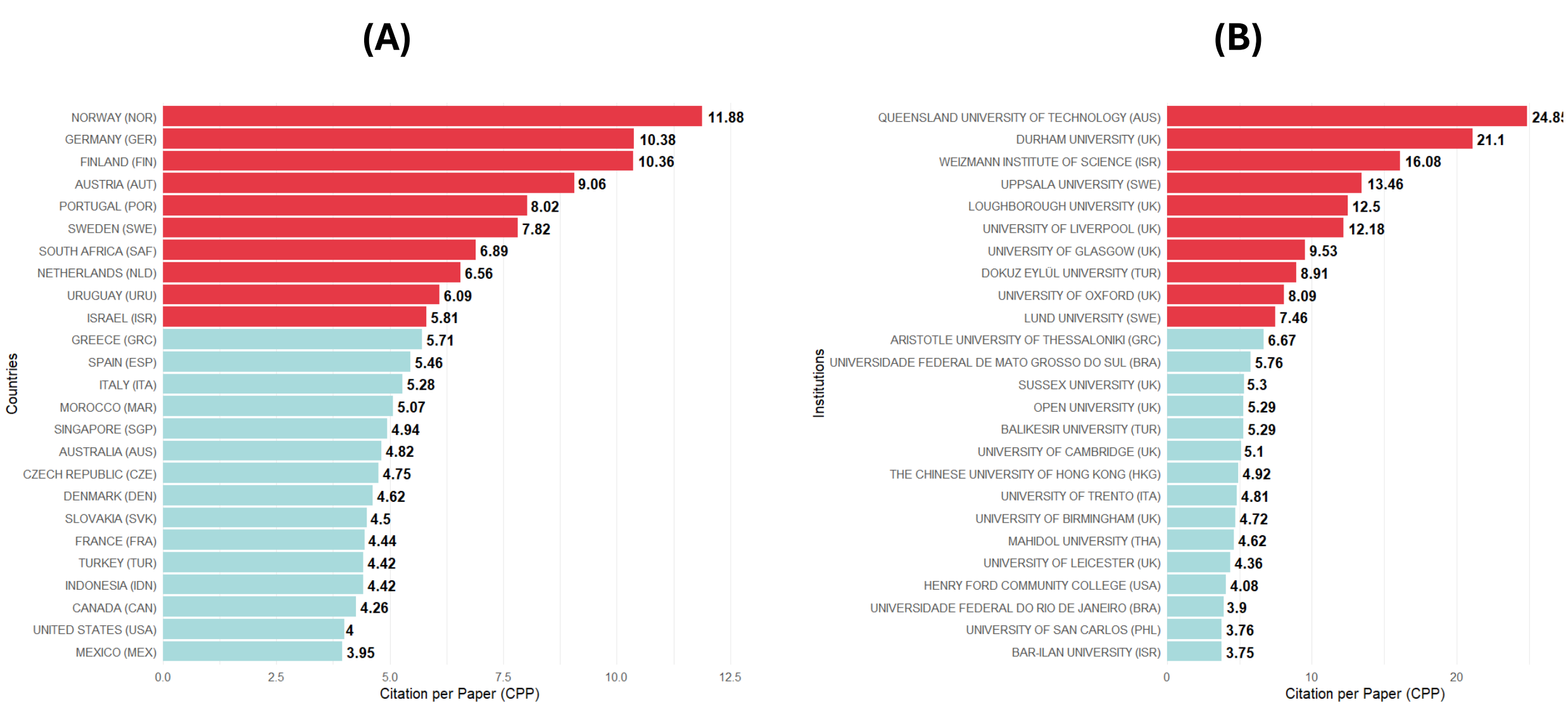

**Fig. 6** Leading Countries (A) and Institutions (B) by Citation per Paper (CPP) in *Physics Education*

At the institutional level, Queensland University of Technology achieved the highest CPP (24.85), followed by Durham University (21.10), the Weizmann Institute of Science (16.08), and Uppsala University (14.44). Several institutions from the United Kingdom, including Loughborough University (12.50), the University of Liverpool (12.18), the University of Oxford (8.09), and the University of Cambridge (5.10), also ranked among the most influential institutions. The substantial differences in CPP among institutions indicate that research influence was concentrated in a relatively small group of organizations whose publications received exceptionally high levels of scholarly attention.

Fig. 7 presents the country- and institution-level h-index rankings in *Physics Education*. Compared with total citations and citations per paper previously, the h-index provides a more balanced measure of scientific influence by combining publication productivity and citation performance. At the country level, the United Kingdom achieved the highest h-index ($h$ = 28), followed by the United States ($h$ = 20), Germany ($h$ = 19), Italy ($h$ = 17), and Australia ($h$ = 16). These findings indicate that the United Kingdom not only produced the largest number of publications but also generated the greatest number of consistently cited papers within the journal. Germany ranked third despite having substantially fewer publications than the United Kingdom and the United States, highlighting the strong impact of German contributions to PER field.

Several countries with relatively moderate publication outputs also demonstrated strong h-index performance. Sweden achieved an h-index of 14, while Turkey, Spain, and Portugal each recorded an h-index of 12. In addition, Brazil, Israel, and the Netherlands attained h-index values of 11, indicating that their contributions were consistently cited despite producing fewer publications than the leading countries. The presence of Singapore ($h$ = 10), Indonesia ($h$ = 10), India ($h$ = 9), Thailand ($h$ = 7), and Hong Kong ($h$ = 7) among the most influential countries further reflects the growing impact of physics education research originating from emerging research communities in Asia.

At the institutional level, h-index values were concentrated within a relatively small group of universities and research organizations. The Weizmann Institute of Science, University of Oslo, and Lund University achieved the highest h-index values (h = 8), indicating the strongest combination of productivity and citation impact among institutions contributing to *Physics Education*. Several other institutions obtained an h-index of 7, including Uppsala University, University of Western Australia, University of Sydney, University of Cambridge, University of Birmingham, University College London, King's College London, Queensland University of Technology, and Institut Teknologi Bandung. These results suggest that influential contributions to *Physics Education* are distributed across institutions located in Europe, Australia, and Asia rather than being concentrated within a single geographic region.

The h-index rankings also differ noticeably from the total citation rankings. For example, Queensland University of Technology ranked first in total citations but achieved an h-index of 7, indicating that a relatively small number of highly cited papers contributed substantially to its citation performance. In contrast, institutions such as the Weizmann Institute of Science, University of Oslo, and Lund University achieved higher h-index values, suggesting a broader portfolio of consistently cited publications. Therefore, the h-index results demonstrate that research influence within *Physics Education* is shaped not only by publication volume or total citations but also by the ability to produce a sustained body of highly cited work in the field.

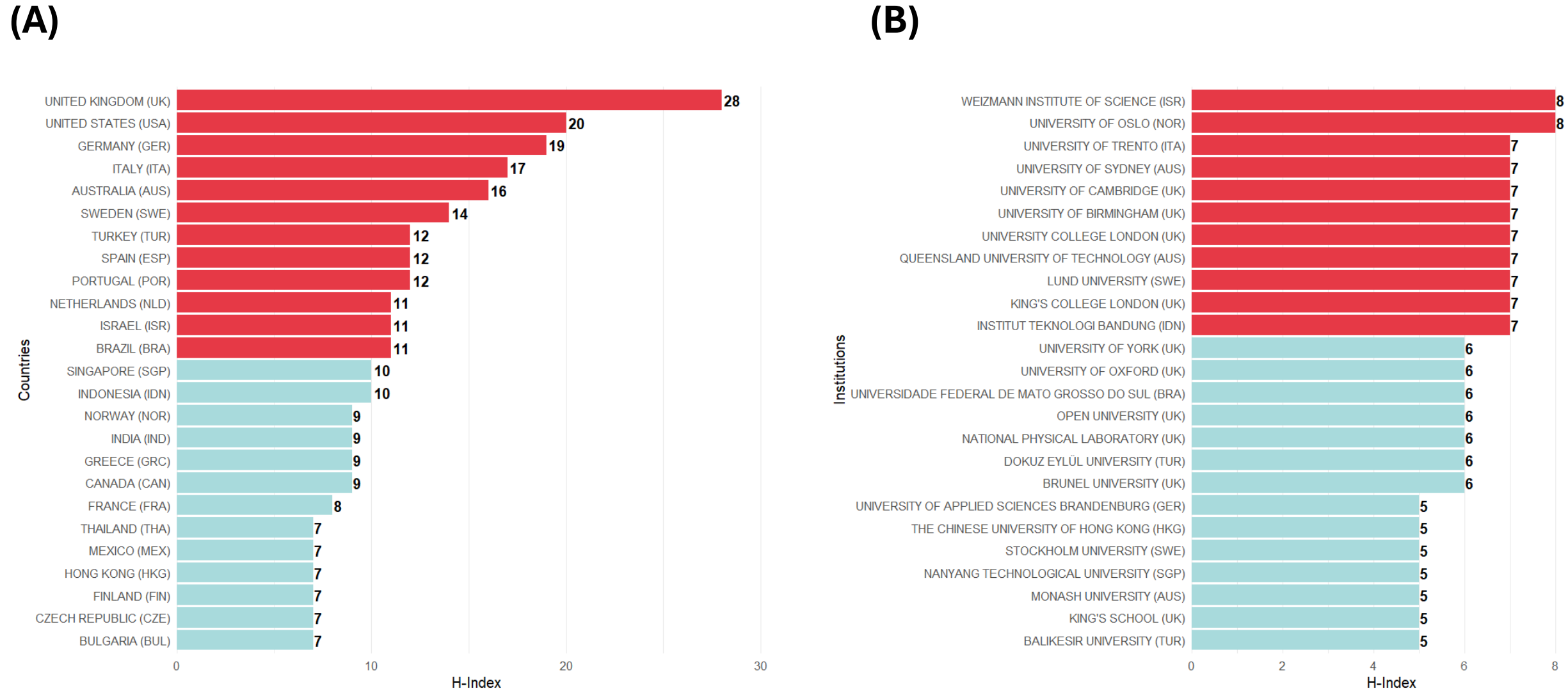

**Fig. 7** The Highest H-Index of Countries (A) and Institutions (B) in *Physics Education*

Collectively, the results demonstrate that publication productivity, total citations, and citation impact (CPP and h-index) represent distinct dimensions of scholarly influence. While the United Kingdom dominated both publication output and total citations, Germany achieved the highest citation impact in terms of citations per paper (CPP), and the United Kingdom attained the highest country-level h-index. Similarly, institutions with the highest CPP were not necessarily those producing the largest numbers of publications, indicating that scholarly influence in *Physics Education* is shaped by both publication volume and citation performance.

Table 1 presents the 15 most cited papers published in *Physics Education*. The first citation pattern indicates that influential research in the journal has been largely shaped by studies addressing student conceptual understanding, misconceptions (Hasan *et al* 1999), and the teaching of fundamental physics concepts. Several highly cited papers focused on persistent learning difficulties in topics such as force (Brown 1989), energy (Goldring and Osborne 1994), thermodynamics, and quantum physics (Ireson 2000, Henriksen *et al* 2014), highlighting the long-standing importance of conceptual change and student reasoning in physics education research. The download counts provide an additional indicator of the reach of these highly cited papers beyond citation-based impact.

**Table 1**.

Top Cited Papers in *Physics Education*

| Authors | Title | Citations | Downloads |
|---|---|---|---|
| Staacks et al (2018) | Advanced tools for smartphone-based experiments: Phyphox | 262 | 19843 |
| Hughes (2005) | Archimedes revisited: A faster, better, cheaper method of accurately measuring the volume of small objects | 245 | 4115 |
| Hasan et al (1999) | Misconceptions and the certainty of response index (CRI) | 180 | 829 |
| Vollmer (2004) | Physics of the microwave oven | 155 | 7638 |
| Yeadon et al (2023) | The death of the short-form physics essay in the coming AI revolution | 135 | 10327 |
| McDermott et al (2000) | Preparing teachers to teach physics and physical science by inquiry | 117 | 1721 |
| Gregorcic and Pendrill (2023) | ChatGPT and the frustrated Socrates | 104 | 13675 |
| Goldring and Osborne (1994) | Students' difficulties with energy and related concepts | 102 | 1391 |
| Williams et al (2003) | Why aren't secondary students interested in physics? | 101 | 1310 |
| Savinainen and Scott (2002) | The force concept inventory: A tool for monitoring student learning | 97 | 4079 |
| Whitelegg and Parry (1999) | Real-life contexts for learning physics: Meanings, issues and practice | 90 | 1555 |
| Henriksen et al (2014) | Relativity, quantum physics and philosophy in the upper secondary curriculum: Challenges, opportunities and proposed approaches | 87 | 2433 |
| Zetie et al (2000) | How does a Mach-Zehnder interferometer work? | 87 | 12937 |
| Brown (1989) | Students' concept of force: The importance of understanding Newton's third law | 87 | 2345 |
| Ireson (2000) | The quantum understanding of pre-university physics students | 83 | 866 |

A second prominent pattern involves innovations in pedagogy and teacher education. Highly cited contributions examined inquiry-based teaching (McDermott *et al* 2000), real-world learning contexts (Whitelegg and Parry 1999), and the use of research-based assessment instruments to monitor student learning (Savinainen and Scott 2002), reflecting the journal's emphasis on improving classroom teaching practice.

More recent highly cited papers also demonstrate increasing interest in educational technologies and digital transformation. Smartphone-based experiments (Staacks *et al* 2018), video-analysis tools (Hughes 2005), and artificial intelligence (Gregorcic and Pendrill 2023, Yeadon *et al* 2023) have emerged as influential topics, indicating a shift toward technology-enhanced learning environments and contemporary educational challenges in physics education research. Overall, the most cited papers suggest that the intellectual development of *Physics Education* has been driven by a combination of experimental physics, technological advancement, conceptual physics, and teachers' practice.

### B. Topic modeling

Considering the balance between topic distinctiveness, interpretability, and model performance, a five-topic solution was selected for reporting the topic modeling results. Table 2 presents the five major research themes identified through latent Dirichlet allocation (LDA) topic modeling of publication abstracts published in *Physics Education* between 1966 and 2026 including representative words and sample papers. The extracted topics reveal the journal's intellectual structure. Altogether, these themes indicate a strong emphasis on experimental physics instruction, conceptual learning, and pedagogical innovation throughout the journal's publication history.

**Table 2.**

Five Extracted Topics of *Physics Education* Between 1966 and 2026

| **Topic** | **Label** | **Representative Keywords** | **Example** |
|---|---|---|---|
| Topic 1 | Force and motion experiments (FME) | *simple, datum, energy, measurement, learn, problem, laboratory, determine, science, current, work, make, force, motion, system, test, value, phenomenon, project, effect* | (Hughes 2005) |
| Topic 2 | Energy and light experiments (ELE) | *approach, experimental, energy, light, time, university, science, course, work, different, make, analysis, project, problem, measure, force, simple, develop, activity, discuss* | (Zetie *et al* 2000) |

| Topic 3 | Laboratory based learning (LBL) | *learn, method, simple, time, activity, system, example, measure, discuss, find, experimental, measurement, law, datum, speed, teacher, work, classroom, technology, tool* | (Staacks *et al* 2018) |
|---|---|---|---|
| Topic 4 | Conceptual understanding (CU) | *concept, understand, design, datum, lean, material, model, field, help, experimental, science, physical, group, laboratory, make, process, find, work, different, develop* | (Brown 1989) |
| Topic 5 | Physics teacher education (PTE) | *teacher, model, year, understand, time, object, method, motion, course, relate, scientific, datum, law, may, work, force, way, analysis, real, effect* | (McDermott *et al* 2000) |

Topic 1 (*force and motion experiments*, FME) is characterized by keywords such as *force*, *motion*, *measurement*, *laboratory*, *determine*, and *apparatus*. This theme focuses on the development and implementation of experiments designed to demonstrate mechanical concepts and support hands-on learning of force and motion phenomena.

Topic 2 (*energy and light experiments*, ELE) contains keywords including *approach, experimental, energy*, *light*, *time, university,* and *science*. The topic reflects studies that employ demonstrations and experimental activities to teach energy-related concepts, optics, and wave phenomena through practical investigation.

Topic 3 (*laboratory-based learning*, LBL) is represented by keywords such as *learn*, *method*, *technology*, *tool*, *simulation*, and *system*. This theme highlights the growing integration of technology, data acquisition systems, and physical measurement tools to enhance experimental learning experiences in physics education classroom.

Topic 4 (*conceptual understanding*, CU) is dominated by terms such as *concept*, *understand*, *model*, *develop*, *process* and *analysis*. The topic reflects research concerned with students' conceptual understanding, conceptual change, and the development of scientific knowledge through learning activities and inquiry processes.

Topic 5 (*physics teacher education*) includes keywords such as *teacher*, *method*, *scientific*, *course*, *application*, *principle*, and *demonstration*. This theme encompasses studies on teaching strategies, curriculum application, classroom practices, and pedagogical approaches aimed at improving the effectiveness of physics instruction.

Overall, the five-topic solution reveals a strong emphasis on experiment-based physics learning, with the first three topics focusing on laboratory activities, experimental design, and technology-enhanced instruction. The remaining two topics highlight the importance of conceptual understanding and instructional innovation, indicating that *Physics Education* has consistently balanced practical experimentation with broader concerns about teaching effectiveness and student learning over the past six decades.

Nevertheless, emerging research areas such as modern physics (Vollmer 2004) and quantum physics education (Henriksen *et al* 2014) and artificial intelligence in physics education (Yeadon *et al* 2023, Gregorcic and Pendrill 2023) did not emerge as distinct topics in the five-topic solution, despite the presence of several highly cited publications addressing these subjects. This finding suggests that, although these topics have attracted considerable scholarly attention and citation impact, their overall publication volume remains relatively small compared with the dominant body of research on laboratory experiments, instructional demonstrations, and hands-on physics learning. Consequently, these areas have not yet accumulated sufficient representation within the corpus to form distinct thematic clusters. Their absence as standalone topics may therefore indicate promising directions for future research and signal emerging areas of interest within the PER community. As publication activity in these areas continues to grow, they may evolve into distinct research themes within the journal.

Fig. 8 presents the temporal prevalence of the five major research themes identified in *Physics Education* between 1966 and 2026. The figure illustrates how the relative prevalence of the extracted topics evolved across five-year publication intervals. Overall, the thematic distribution remained relatively balanced throughout most of the journal's history, indicating the coexistence of several enduring research traditions. Nevertheless, gradual changes can be observed in the relative prominence of the five themes over time. The findings suggest that *Physics Education* has evolved from a journal strongly oriented toward experimental activities to one that increasingly emphasizes conceptual learning and pedagogical development while maintaining its long-standing commitment to laboratory-based instruction.

The thematic evolution of *Physics Education* can be interpreted in three phases. The first phase, experiment-oriented physics education (1966-1986), was characterized by the strong prevalence of physics teacher education (PTE) and energy and light experiments (ELE), alongside substantial contributions from force and motion experiments (FME) and laboratory-based learning (LBL). This pattern suggests that early publications primarily focused on classroom demonstrations, laboratory activities, and instructional approaches designed to support the teaching of fundamental physics concepts.

The second phase, thematic diversification (1986-2006), exhibited relatively balanced prevalence across all five themes. During this period, no single topic consistently dominated the journal, indicating a broadening of research interests beyond traditional laboratory instruction. Themes related to experiments, teaching practices, and conceptual learning coexisted and contributed to a more diversified intellectual structure.

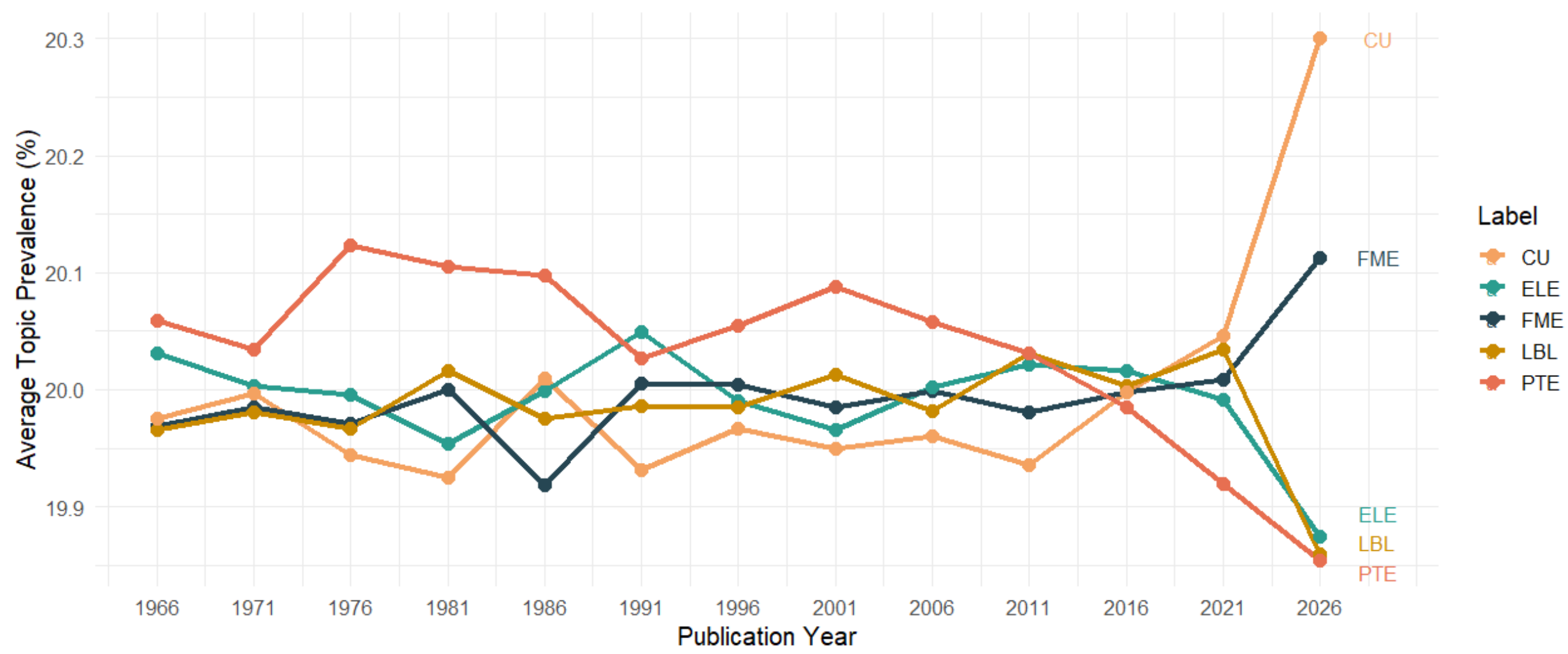

**Fig. 8** Prevalence of Research Topics in *Physics Education* over Six Decades

The third phase, conceptual and pedagogical advancement (2006-2026), was marked by the growing prominence of conceptual understanding (CU). Although all themes maintained relatively similar prevalence levels for much of the period, CU gradually increased and became the most prevalent theme by 2026. In contrast, energy and light experiments (ELE) experienced renewed growth after 2006 and remained highly visible until approximately 2021 before declining in the most recent period. force and motion experiments (FME) displayed remarkable stability throughout the journal's history and even increased slightly in recent years, highlighting the enduring importance of foundational mechanics concepts in physics education research.

The increasing prominence of conceptual understanding reflects the growing interest of PER scholars in conceptual change, scientific reasoning, misconceptions, assessment, and evidence-based approaches to physics learning. This trend suggests a gradual shift in attention from how physics experiments are conducted toward how students develop and apply conceptual understanding of physics. It is also possible that challenges associated with understanding emerging topics, including quantum physics, are partly represented within this broader thematic cluster. However, the number of studies explicitly focused on quantum physics education appears insufficient to form a distinct thematic category. Consequently, conceptual understanding should be interpreted as a broad research area encompassing diverse conceptual issues rather than a theme specifically devoted to quantum physics education.

An interesting finding is that topic 1 (force and motion experiments, FME) never emerged as the most prevalent theme throughout the period under investigation. Nevertheless, its prevalence remained remarkably stable across nearly all publication periods, indicating its enduring importance within the field. This stability suggests that research on force and motion experiments represents a fundamental and persistent component of PER. Given that force and motion constitute core concepts that underpin more advanced areas of physics, studies related to their experimental investigation have remained relevant across generations of researchers, even as newer themes gained prominence. Furthermore, an equally interesting trend concerns physics teacher education (topic 5). Although this theme maintained a relatively high prevalence throughout much of the journal's history, its prominence gradually declined after 2006. This decline does not necessarily indicate reduced importance. Rather, it may suggest that issues related to teacher development and classroom practice have become increasingly integrated into broader educational themes such as conceptual understanding, assessment, and laboratory-based learning. Overall, the thematic analysis indicates that experimental and laboratory-based instruction remains a defining characteristic of *Physics Education*. More recently, however, the attention of PER scholars has gradually shifted from the development of physics experiments toward improving students' conceptual understanding and enhancing the effectiveness of physics teaching and learning.

### 4. Discussion and Recommendation

This study examined the development of *Physics Education* between 1966 and 2026 through bibliometric and topic modeling analyses. The results revealed a substantial increase in publication productivity over six decades, accompanied by the growing internationalization of research contributions. While a small number of countries and institutions dominated publication output, the citation analyses indicate that productivity and scholarly influence do not necessarily coincide. Different impact measures revealed complementary perspectives on research performance, highlighting the multifaceted nature of scientific influence within the journal. Topic modeling identified five major research themes that constitute the intellectual structure of the journal. Collectively, these themes demonstrate the journal's long-standing emphasis on experiment-based instruction, laboratory activities, and the improvement of physics teaching and learning. The thematic evolution analysis further revealed a gradual transition from traditional laboratory demonstrations and experimental apparatus design toward conceptual understanding, student learning processes, and pedagogical innovation.

These findings are partly consistent with previous reviews of physics education research. Odden et al (2020) highlighted the prominence of topics such as representations, problem solving, laboratory instruction, conceptual assessment, and both qualitative and quantitative research approaches in PER. Similarly, Yun (2020) identified introductory physics, force and motion, and pedagogical content knowledge (PCK) as recurring areas within the field. Santoso et al (2022) also emphasized the importance of critical thinking, science process skills, and scientific literacy in the twenty-first-century physics education. The present study confirms the enduring relevance of several of these themes, particularly force and motion experiments, conceptual understanding, and physics teaching. However, it also reveals a distinctive thematic pattern within a journal, highlighting the long-standing prominence of experimental designs, laboratory-based teaching, and technology-supported learning over time.

However, several notable differences emerged. While previous reviews by Odden et al (2020) and Yun (2020) identified broad areas of research within physics education, the present analysis revealed more content-specific thematic structures within *Physics Education*. In particular, experimental research was represented by two distinct themes, namely force and motion experiments (FME) and energy and light experiments (ELE), reflecting the journal's long-standing emphasis on physics-content-based experimental designs. Furthermore, technology-supported experimental learning emerged as a separate theme, suggesting the increasing integration of sensors, simulations, data-acquisition systems, and digital tools into laboratory instruction. Interestingly, topics such as quantum physics education, artificial intelligence, machine learning, and learning analytics did not emerge as independent thematic clusters despite the presence of highly cited papers in these areas. This suggests that although these topics have attracted considerable scholarly attention, their publication volume remains insufficient to form standalone themes within the journal's dominant experimental and pedagogical traditions.

Several limitations should be acknowledged. First, the study was restricted to publications indexed in a single journal, which may not fully represent the broader landscape of physics education research. Second, approximately 18% of the initially retrieved records were excluded during screening, particularly because of missing affiliation information, which may have limited country- and institution-level analyses. Third, the impact analysis relied primarily on citation-based indicators, which may not fully capture the educational influence of publications among physics teachers and classroom practitioners. Future studies are encouraged to complement citation metrics with article download statistics and other usage-based indicators to provide a broader assessment of how research is accessed and utilized within the physics education community. Fourth, the topic modeling analysis was conducted using publication abstracts rather than full-text documents. Although abstracts have generally captured the main objectives, methods, and findings of a study, they may not fully represent the depth and complexity of the underlying research. Consequently, some thematic nuances and emerging topics may not have been adequately captured by the LDA model. This limitation was primarily due to restricted access to the complete collection of published articles. Future studies are therefore encouraged to extend the present analysis by utilizing full-text documents, which may provide a more comprehensive understanding of the intellectual structure, thematic evolution, and emerging research directions within Physics Education. Finally, the topic modeling results are influenced by preprocessing decisions, topic-number selection, and the assumptions of the LDA algorithm. Different modeling approaches may therefore yield alternative thematic structures.

Future studies could expand the scope of analysis by incorporating multiple journals, conference proceedings, and additional databases such as Web of Science or Dimensions. Comparative investigations across publication venues may offer a broader understanding of the evolution of physics education research worldwide. Researchers may also employ more advanced topic-modeling techniques, including BERTopic, dynamic topic modeling, and large language model-based approaches, to better capture thematic changes over time. Particular attention should be given to emerging areas such as quantum physics education, artificial intelligence, machine learning, learning analytics, virtual laboratories, and computational physics education, which appear to represent promising future directions for the physics education community. In conclusion, *Physics*

*Education* has evolved from a journal primarily focused on laboratory experimentation and instructional demonstrations into a broader platform emphasizing conceptual understanding, pedagogical innovation, and technology-enhanced learning. The findings provide valuable insights into the historical development of the journal and offer evidence-based guidance for future research in physics education.

## Data and code availability

The bibliometric dataset and R codes run in the present study have been made available in GitHub repository (https://github.com/santosoph/PED-Bibliometric ).

## References

Aria M and Cuccurullo C 2017 bibliometrix: An R-tool for comprehensive science mapping analysis *J. Informetr.* **11** 959–75

Blei D M, Ng A Y and Jordan M I 2003 Latent Dirichlet allocation *Journal of Machine Learning Research* **3** 993–1022

Brown D E 1989 Students' concept of force: The importance of understanding Newton's third law *Phys. Educ.* **24** 353

Campbell J, Ansell K and Stelzer T 2024 Evaluating IBM's Watson natural language processing artificial intelligence as a short-answer categorization tool for physics education research *Phys. Rev. Phys. Educ. Res.* **20** 010116

Chen Z, Xu M, Garrido G and Guthrie M W 2020 Relationship between students' online learning behavior and course performance: What contextual information matters? *Phys. Rev. Phys. Educ. Res.* **16** 010138

Docktor J L and Mestre J P 2014 Synthesis of discipline-based education research in physics *Physical Review Special Topics - Physics Education Research* **10** 020119

Goldring H and Osborne J 1994 Students' difficulties with energy and related concepts *Phys. Educ.* **29** 26

Gregorcic B and Pendrill A M 2023 ChatGPT and the frustrated Socrates *Phys. Educ.* **58** 035021

Hasan S, Bagayoko D and Kelley E L 1999 Misconceptions and the Certainty of Response Index (CRI) *Phys. Educ.* **34** 294

Henriksen E K, Bungum B, Angell C, Tellefsen C W, Frågåt T and Bøe M V 2014 Relativity, quantum physics and philosophy in the upper secondary curriculum: Challenges, opportunities and proposed approaches *Phys. Educ.* **49** 678

Hughes S W 2005 Archimedes revisited: A faster, better, cheaper method of accurately measuring the volume of small objects *Phys. Educ.* **40** 468

Ireson G 2000 The quantum understanding of pre-university physics students *Phys. Educ.* **35** 15

Kaps A, Splith T and Stallmach F 2021 Implementation of smartphone-based experimental exercises for physics courses at universities *Phys. Educ.* **56** 035004

Kaps A and Stallmach F 2022 Development and didactic analysis of smartphone-based experimental exercises for the smart physics lab *Phys. Educ.* **57** 045038

Liu Y, Lei Q, He X, Xue Y, He K, Yang H, Wang Y, Zhang X, Yang L, Zhou Y, Hu R and Xie Y 2025 Building an affordable self-driving lab: Practical machine learning experiments for physics education using Internet-of-Things *APL Machine Learning* **3** 046105

McDermott L C and Redish E F 1999 Resource Letter: PER-1: Physics Education Research *Am. J. Phys.* **67** 755–67

McDermott L C, Shaffer P S and Constantinou C P 2000 Preparing teachers to teach physics and physical science by inquiry *Phys. Educ.* **35** 411

Odden T O B, Marin A and Caballero M D 2020 Thematic analysis of 18 years of physics education research conference proceedings using natural language processing *Phys. Rev. Phys. Educ. Res.* **16** 010142

Organtini G and Tufino E 2022 Effectiveness of a Laboratory Course with Arduino and Smartphones *Educ. Sci. (Basel).* **12** 898

Pace J, Hansen J and Stewart J 2024 Exploring techniques to improve machine learning's identification of at-risk students in physics classes *Phys. Rev. Phys. Educ. Res.* **20** 010149

Röder M, Both A and Hinneburg A 2015 Exploring the space of topic coherence measures *WSDM 2015 - Proceedings of the 8th ACM International Conference on Web Search and Data Mining* pp 399–408

Santoso P H, Istiyono E, Haryanto and Hidayatulloh W 2022 Thematic Analysis of Indonesian Physics Education Research Literature Using Machine Learning *Data (Basel).* **7** 147

Savinainen A and Scott P 2002 The force concept inventory: A tool for monitoring student learning *Phys. Educ.* **37** 45

Staacks S, Hütz S, Heinke H and Stampfer C 2018 Advanced tools for smartphone-based experiments: Phyphox *Phys. Educ.* **53** 045009

Vollmer M 2004 Physics of the microwave oven *Phys. Educ.* **39** 74

Whitelegg E and Parry M 1999 Real-life contexts for learning physics: Meanings, issues and practice *Phys. Educ.* **34** 68

Williams C, Stanisstreet M, Spall K, Boyes E and Dickson D 2003 Why aren't secondary students interested in physics? *Phys. Educ.* **38** 324

Yeadon W, Inyang O O, Mizouri A, Peach A and Testrow C P 2023 The death of the short-form physics essay in the coming AI revolution *Phys. Educ.* **58** 035027

Yun E 2020 Review of trends in physics education research using topic modeling *Journal of Baltic Science Education* **19** 388–400

Zetie K P, Adams S F and Tocknell R M 2000 How does a Mach-Zehnder interferometer work? *Phys. Educ.* **35** 46

Zhao Y 2026 Smartphone-based undergraduate physics labs: a comprehensive review of innovation, accessibility, and pedagogical impact *Eur. J. Phys.* **47** 013001